# On the Convex Geometry of Weighted Nuclear Norm Minimization

Seyedroohollah Hosseini

roohollah.hosseini@outlook.com

Abstract- Low-rank matrix approximation, which aims to construct a low-rank matrix from an observation, has received much attention recently. An efficient method to solve this problem is to convert the problem of rank minimization into a nuclear norm minimization problem. However, soft-thresholding of singular values leads to the elimination of important information about the sensed matrix. Weighted nuclear norm minimization (WNNM) has been proposed, where the singular values are assigned different weights, in order to treat singular values differently. In this paper the solution for WNNM is analyzed under a particular weighting condition using the connection between convex geometry and compressed sensing algorithms. It is shown that the WNNM is convex where the weights are in non-descending order and there is a unique global minimizer for the minimization problem.

## 1. Introduction

Low rank matrix approximation, which is the recovery of an unknown low rank matrix from very limited number of known entries, has gained growing interest in recent years. It has wide variety of applications in computer vision and machine learning. For example, human facial images can construct a low rank nature of matrix that is able to recover corrupted or occluded faces [3], [6], [13]. Another example of the application of low rank matrix approximation is in control [7]. Regarding fast development of convex and non-convex optimization techniques in recent years, a considerable amount of improved algorithms and modified models have been proposed for low rank matrix approximation [8], [10].

Recently, some researchers have focused on nuclear norm minimization in order to approximate a low rank matrix. A low rank matrix $X$ from observation $Y$ can be recovered as the solution of the following objective function

$$\hat{X} = \arg \min \|Y - X\|_F^2 + \lambda \|X\|_*$$

where $\lambda$ is a positive constant which balances the loss function and the nuclear norm. In general, nuclear norm is defined as follows:

$$\|A\|_* = tr\left(\sqrt{A^*A}\right) = \sum_{i=1}^{min\{m,n\}} \sigma_i(A) \tag{1}$$

The nuclear norm minimization of matrix $A$ with $m$ rows and $n$ columns aims to minimize (1), i.e.

$$min \sum_{i=1}^{\min\{m,n\}} \sigma_i(A) \qquad (2)$$

where $\sigma_i$ is the i-th singular value.

The equation (2) treats all singular values democratically and shrinks each singular value equally. However, larger singular values carry more important information than smaller ones. Moreover, in many cases we have prior information about the singular values. For instance, diffuse optical tomography (DOT) incorporates prior knowledge to build a new image reconstruction algorithm [12]. Thus, soft thresholding operator of nuclear norm (2) fails to approximate a low rank matrix regarding such understanding of the problem.

Gu [4] proposed weighted nuclear norm minimization in order to add more flexibility to (2). By assigning a particular weight to each singular value, weighted nuclear norm minimization becomes a new minimization method in which larger singular values are penalized less. This is the main idea that explains why many researchers proposed the usage of weights for minimization problems such as $\ell_1$ norm minimization, reducing smaller elements in the minimization than larger ones. The weighted nuclear norm is defined as follows:

$$\|A\|_*^w = \sum_i w_i \sigma_i(A) \qquad (3)$$

Various weighting algorithms can be developed based on the prior knowledge and understanding of the problem for solving (3). But, before choosing each weight for assigning to corresponding singular value it is vital to realize what orders of weights could lead to find the solution for the problem. We prove that if the weights are in non-decreasing order, $w_1 \leq w_2 \leq \cdots \leq w_n$ , then the minimization for equation (3) has a global optimal solution.

It has known that there is a correlation between the optimality condition for linear inverse problems and the descent cone of the proper convex function. Amelunxen [1] have proved that a linear inverse problem with random measurements succeeds with high probability when the number of measurements surpasses the statistical dimension of the proper convex cone. They defined an algorithm for studying the statistical dimension of a descent cone by converting questions about the statistical dimension of a descent cone into question about subdifferential. The algorithm works based on the computation of distance between the subdifferential of the proper convex function and a Gaussian matrix $N(0, I)$. We employ the same method to identify the minimum amount of measurements for weighted nuclear norm minimization. We prove that under a certain condition it is possible to identify how many measurements are needed to find the unique solution for the minimization problem.

In what follows, the contribution of this paper will be alluded to briefly. First, we bring up the definition of a descent cone. We also mention necessary concepts for the statistical dimension a descent cone. Then, we point out a theorem which clarifies the least number of measurements to solve a linear inverse problem. In the end, the statistical dimension of weighted nuclear norm will be calculated and the existence of a unique minimizer for the calculated statistical dimension will be proved under a certain condition.

The notations in the paper are standard. However, we notice few notations which is thought to bring ambiguity for the readers. Other notations will be clarified in the section 2.

We assume the rank of the processing matrix $A$ is $r \leq n$ where $w_j = 0$ for $j > r$. We denote the set of weights for non-zero singular values by $T = \{w_1, w_2, \dots, w_r\}$. Moreover, we will benefit of employing Frobenius norm which can be defined in different ways.

$$\|A\|_F = \sqrt{\sum_{i=1}^{m} \sum_{j=1}^{n} |a_{ij}|^2} = \sqrt{trace\ (A^*A)} = \sqrt{\sum_{i=1}^{\min\{m,n\}} \sigma_i^2}$$

## 2.  Phase transition:

Let us begin this section by stating few preliminaries for the problem. It is worth to note that for the equation (3), we assume the order of the weights are non-descending because we wish to dwindle larger singular values less. Furthermore, the weights are normalized and bounded between zero and one. So, we have

$$1 \leq w_1 \leq w_2 \leq \dots \leq w_n \leq 0$$

where $w_i$ is the weight assigned to $\sigma_i$. Using this order of weights, we prove that the weighted nuclear norm minimization has a solution. Moreover, it is well known that a linear inverse problem has a unique optimal point $(x_0)$ if and only if

$$\mathcal{D}(f, x_0) \cap ker\ (A) = \{0\}$$

where $\mathcal{D}(f, x_0)$ is the descent cone of the problem at point $x_0$ and $ker\ (A)$ denotes null-space property for the regularizer $A$. This is one of the most popular method of using the descent cone of the proper function to investigate if the linear inverse problem has a solution. However, we do not employ it in this paper.

### 2.1. Statistical Dimension:

We begin by bringing up few concepts related to descent cone. Then, the definition of the statistical dimension of a descent cone will be mentioned. Afterwards, the correlation between the statistical dimension and phase transition will be pointed out. Finally proposition 1 will show how to calculate the statistical dimension for weighted nuclear norm using subdifferential.

Definition 1 (Descent cone) [1, definition 2.7]: The descent cone of a proper convex function at the point $x_0$ is a set of functions which do not increase the function near the point $x_0$.

$$\mathcal{D}(f, x_0) = \{z \ \epsilon \ \mathbb{R}^d \mid \exists \ \tau > 0 \colon f(x_0 + \tau z) \leq f(x_0)\}$$

Definition 2 (Statistical dimension) [1, proposition 2.4]: If $C \in \mathbb{R}^d$ is a closed convex cone, then the statistical dimension of $C$ is denoted by $\delta(C)$, i.e.

$$\delta(C) = \mathbb{E}[\|\textstyle\prod_C(\mathcal{G})\|] \quad \text{where } \mathcal{G} \sim \text{NORMAL}(0, I)$$

where $\prod_C(\mathcal{G})$ denotes the Euclidean projection onto the cone $C$, that is

$$\textstyle\prod_C(\mathcal{G}) = argmin_{c \epsilon C} \|c - x\|_2$$

Later in definition 3, we will define Euclidean projection onto sets and use it to prove nuclear norm minimization has a unique solution.

Now we state a theorem for phase transition in linear inverse problems with random measurements. This theorem provides a threshold number of measurements for the success of the linear inverse problem. It utilizes the statistical dimension of a descent cone which is defined in definition 2.

Theorem 1 (necessary condition for a linear inverse problem to succeed) [1, theorem II]: let $m$ be the number of measurements for a linear inverse problem. The problem succeeds with high probability if

$$m \geq \delta(\mathcal{D}(f, x_0)) + \text{ß}\sqrt{d}$$

and fails with high probability if

$$m \leq \delta(\mathcal{D}(f, x_0)) + \text{ß}\sqrt{d}$$

where $x \epsilon \mathbb{R}^d$ and ß is a function that eats a tolerance number between zero and one and spits out a real number.

Theorem 1 tells us that phase transition occurs where the number of measurements becomes greater than the statistical dimension of the descent cone. So, we can conclude that, under minimal assumptions, in order to find the least number of measurements for a linear inverse problem where it is successful it is vital to compute the statistical dimension of the descent cone. The authors of [1] presented a new method for the calculation of $\delta(\mathcal{D}(\|.\|_1, x))$. We use the proposed technique to calculate the statistical dimension of the descent cone for weighted nuclear norm at point $\sigma$, $\delta(\mathcal{D}(\|.\|_*^w), \sigma)$.

### 2.2. Threshold amount of measurements for weighted nuclear norm:

In order to calculate the statistical dimension of the descent cone for WNNM $\delta(\mathcal{D}(\|.\|_*^w), x)$, we need to use subdifferential concept. We initiate this section by mentioning the subdifferential for nuclear norm minimization. Next, we define a function $J(\tau)$ which is expected distance between Gaussian matrix NORMAL$(0, I_d)$ and the calculated subdifferential.

Proposition 1 (Upper bound for the statistical dimension of a descent cone) [1, proposition 4.1]: The statistical dimension of a descent cone, where $f$ is a proper convex function, has the upper bound

$$\delta(\mathcal{D}(f, x)) \leq \inf_{\tau \geq 0} E\left(dist^2(\mathcal{G}, \tau.\partial f(x))\right)$$

Where $\tau$ acts for dilation of the subdifferential in the expression $\tau.\partial f(x)$.

Similar to the proposition 1, we define the statistical dimension of weighted nuclear norm minimization, i.e.

$$\delta(\mathcal{D}(\|.\|_*^w), \sigma) \leq \inf_{\tau \geq 0} E\left(dist^2(\mathcal{G}, \tau.\partial\|X\|_*^w)\right) \tag{4}$$

where the subdifferential $\partial\|X\|_*^w$ is nonempty, compact, and does not contain the origin.

we denote the right-hand side of the inequality (4) by

$$J(\tau) = E\left(dist^2(\mathcal{G}, \tau.\partial\|X\|_*^w)\right) \quad \text{for } \tau > 0$$

Now, we calculate $J(\tau)$ and demonstrate there is a unique global minimizer for it. As we mentioned, the minimizer of $J(\tau)$ supplies the minimum number of measurements needed for successful weighted nuclear norm minimization.

we commence the calculation by considering fixed-dimension setting. Assume that there is a matrix $X$ which is $m \times n$ ($m \leq n$) and has rank $r$, i.e.

$$X = \begin{bmatrix} \Sigma & 0 \\ 0 & 0 \end{bmatrix} \quad \text{where } \Sigma = diag(\sigma_1, \sigma_2, \dots, \sigma_r) \quad \text{and } \sigma_i \geq 0 \text{ for } i = 1, \dots, r.$$

Similarly, we partitioned matrix $G$ which conform $X$ :

$$G = \begin{bmatrix} G_{11} & G_{12} \\ G_{21} & G_{22} \end{bmatrix} \quad \text{where } G_{11} \text{ is } r \times r \quad \text{and} \quad G_{22} \text{ is } (m-r) \times (n-r)$$

The subdifferential of nuclear norm is mentioned in [11, example 2]. Since the subdifferential of the weighted nuclear norm is only depends on the sign pattern of $w_i$, and $w_i$ has the same sign pattern to $X$, can calculate the subdifferential of the nuclear norm similarly. We are going to discuss about it later in lemma 1. The subdifferential of the nuclear norm at $X$ has the following form:

$$\partial\|.\|_*^w = \left\{ \begin{bmatrix} W_r & 0 \\ 0 & W_n K \end{bmatrix} : \sigma_1(K) \leq 1 \right\} \quad \text{where } \sigma_1(K) \text{ is the largest singular value,}$$

$$W_r = diag(w_1, \dots, w_r), \ W_n = diag(w_{r+1}, w_{r+2}, \dots, w_n) \text{ and } 0 \leq w_1 \leq w_2 \leq \dots \leq w_n \leq 1.$$

Now we calculate the distance function.

$$dist^2(G, \tau.\partial\|X\|_*^w) = \left\| \begin{bmatrix} G_{11} - \tau W_r & G_{12} \\ G_{21} & 0 \end{bmatrix} \right\|_F^2 + \inf_{\substack{\sigma_1(K) \leq 1 \\ j \notin T}} \left\| G_{22} - \tau w_j K \right\|_F^2 \tag{5}$$

In order to solve the right-hand side of equation (5), Wielandt- Hoffman theorem is exploited which defines a lower bound for Frobenius norm using singular values.

Theorem 2 (Wielandt- Hoffman Theorem) [5, corollary 7.3.8]: let $A$ and $B$ are normal matrices and $C = A - B$. there is a lower bound for C, i.e.

$$\sum_{i=1}^{n} |a_i - b_i|^2 \leq \|C\|_F^2$$

Where $a_i$ and $b_i$ are the eigenvalues of $A$ and $B$ respectively such that $\sum_{i=1}^{n} |a_i - b_i|^2$ is the minimum for all possible orderings.

Using theorem 2 and equation (5) we derive

$$\inf_{\substack{\sigma_1(K) \leq 1 \\ j \notin T}} \left\| G_{22} - \tau w_j K \right\|_F^2 = \inf_{\substack{\sigma_1(K) \leq 1 \\ j \notin T}} \sum_{i=1}^{m-r} (\sigma_i(G_{22}) - \tau w_j \sigma_i(K))^2$$

where $\sigma_i(.)$ is the i-th largest singular values. Because $\sigma_i(K) \leq 1$ we can write

$$\inf_{\substack{\sigma_1(K) \\ j \notin T}} \left\| G_{22} - \tau w_j K \right\|_F^2 = \inf_{\substack{\sigma_1(K) \\ j \notin T}} \sum_{i=1}^{m-r} (\sigma_i(G_{22}) - \tau w_j \sigma_i(K))^2 = \sum_{\substack{i=1 \\ j \notin T}}^{m-r} Pos^2 (\sigma_i(G_{22}) - \tau w_j) \qquad (6)$$

where $Pos(x) = x \vee 0$ and operator $\vee$ returns the maximum of two values. Taking the expectation of (6) while bearing in mind that the expectation of a sum is the sum of the expectation,

$$\text{E}(\text{dis}^2\,(\mathcal{G}\,,\tau\,.\,\partial(\|X\|_*,\text{w}\,)) = \sum_{j \in T} \text{r}(m + n - r + \tau^2 w_j^2) + \text{E}\left[\sum_{\substack{i=1 \\ j \notin T}}^{m-r} Pos^2 (\sigma_i(G_{22}) - \tau w_j)\right] \qquad (7)$$

To solve expectation in (7) we utilize a theorem for the joint density function of Wilshart distribution. But before that we bring another theorem and justify why $\sigma_i(G_{22})$ can be found by calculation $\sigma_i(B)$ where $B$ has Wilshart distribution.

Theorem 3 (Wilshart distribution density) [2, theorem 7.2.2]: if $Z_1, Z_2, \ldots, Z_n$ are independently distributed Normal variables $N(0, \Sigma)$, the density of $A = \sum_{\alpha=1}^{n} Z_\alpha Z'_\alpha$ is denoted by $W(A|\Sigma, n)$ and, where $A$ is positive definite, equals to

$$\frac{|A|^{\frac{1}{2}(n-p-1)} e^{-\frac{1}{2} Tr(\Sigma^{-1}A)}}{2^{\frac{1}{2}pn}\, \Pi^{\frac{p(p-1)}{4}}\, |\Sigma|^{\frac{1}{2}n}\, \prod_{i=1}^{p} \Gamma \frac{1}{2}(n+1-i)}$$

where $\Gamma$ is the multivariate Gamma function, i.e.

$$\Gamma_p(t) = \pi^{\frac{p(p-1)}{4}} \prod_{i=1}^{p} \Gamma[t - \frac{1}{2}(i-1)]$$

Because the singular values for matrix A has the following definition

$$\sigma_i(G_{22}) = \sqrt{\lambda_i(\acute{G}_{22}G_{22})}$$

it suffices to calculate eigenvalues of $G_{22}'G_{22}$. The singular values are positive. Moreover, $G_{22}$ has Normal distribution of $N(0, \Sigma)$. Therefore, $G_{22}'G_{22}$ has Wilshart distribution. Using the next theorem, we find the density of the eigenvalues of $G_{22}'G_{22}$.

Theorem 4 (Density of eigenvalues of Wilshart distribution) [2, theorem 13.3.2]: the density of eigenvalues $(\lambda_i)$ for $A_{p \times p}$, which has Wilshart distribution W(I,n), where $(\lambda_1 \geq \lambda_2 \geq ... \geq \lambda_n)$ is equal to

$$\frac{\pi^{\frac{1}{2}p^2} \prod_{i=1}^{p} \lambda_i^{\frac{1}{2}(n-p-1)} e^{\frac{-1}{2}\sum_{i=1}^{p}\lambda_i} \prod_{i<j}(\lambda_i - \lambda_j)}{2^{\frac{1}{2}pn}\Gamma_p(\frac{1}{2}n)\Gamma_p(\frac{1}{2}p)}$$

Using theorem 4 we are able to calculate $E\left[\sum_{i=1}^{m-r} Pos^2\left(\sigma_i\left(G_{22}\right) - \tau\right)\right]$.

$$E\left[\sum_{\substack{i=1 \\ j \notin T}}^{m-r} Pos^2\left(\sigma_i\left(G_{22}\right) - \tau w_j\right)\right] = \sum_{\substack{i=1 \\ j \notin T}}^{m-r} \int_{\tau w_j}^{\infty} \left(X - \tau w_j\right)^2 \frac{\pi^{\frac{1}{2}(m-r)^2} \prod_{i=1}^{m-r} x_i^{(n-m+r-1)} e^{\frac{-1}{2}\sum_{i=1}^{m-r} x_i^2} \prod_{i<j}(x_i^2 - x_j^2)}{2^{\frac{1}{2}(m-r)n}\Gamma_{m-r}(\frac{1}{2}n)\Gamma_{m-r}(\frac{1}{2}(m-r))} dX \qquad (8)$$

If we put equation (8) into equation (7), we reach the calculated $J(\tau)$, i.e.

$$J(\tau) = r(m + n - r + \tau^2 w_j^2)$$
$$+ (m-r)\int_{\tau w_j}^{\infty}\left(X - \tau w_j\right)^2 \frac{\pi^{\frac{1}{2}(m-r)^2} \prod_{i=1}^{m-r} x_i^{(n-m+r-1)} e^{\frac{-1}{2}\sum_{i=1}^{m-r} x_i^2} \prod_{i<j}(x_i^2 - x_j^2)}{2^{\frac{1}{2}(m-r)n}\Gamma_{m-r}(\frac{1}{2}n)\Gamma_{m-r}(\frac{1}{2}(m-r))} dX$$

Now, we only need to prove that the function $J(\tau) = E\left(dist^2(\mathcal{G}, \tau.\partial\|X\|_*^w)\right)$ has a unique global minimizer for $\tau > 0$. $J(\tau)$ has a unique optimal minimizer if and only if it is continuous and convex. In order to demonstrate the existence of single minimizer for $J(\tau)$, we employ Euclidean projection onto sets. Then, using the correlation between Euclidean projection and distance function, the existence of the unique minimizer for $J(\tau)$ will be shown. But, first we need to justify the convexity for nuclear norm. Now let us begin by pointing out a lemma to show the equality between the subdifferential for weighted nuclear norm and the subdifferential for nuclear norm.

Lemma 1: the subdifferential for nuclear norm and weighted nuclear norm are equal.

it is known that only sign pattern of $\sigma$ can change the result of subdifferential $\partial\|.\|_*$. Since the weights $(w_i)$ and $\sigma$ have the same sign pattern, $\partial\|\sigma\|_*^w = \partial\|\sigma\|_*$.

Before proving the function distance has unique minimizer, we have to show that nuclear norm is convex. Later, we use the fact that nuclear norm is convex to prove the existence of unique minimizer for distance function.

Convexity (nuclear norm): It is sufficient to prove that the nuclear norm is, in fact, a norm. we are to verify the three following requirements. The correctness of these requirements will justify nuclear norm is a norm, and hence nuclear norm is convex.

I.    $\|A\| = 0 \Rightarrow A = 0$
II.   $\|tA\| = |t|\|A\|$
III.  $\|A + B\| \leq \|A\| + \|B\|$

It's trivial to verify the requirements **I** and **II**. The one non-trivial requirement is that the norm satisfies the triangle inequality **III**. To do that, we're going to prove:

$$\sup_{\sigma_1(Q)\leq 1} \langle Q, A\rangle = \sup_{\sigma_1(Q)\leq 1} Tr(Q^H A) = \sum_i \sigma_i\,(A)\ = \parallel A \parallel.$$

Because $\sigma_1(\cdot)$ is itself a norm, what we're actually proving here is that the nuclear norm is dual to the spectral norm.

Let $A = U\Sigma V^H = \sum_i \sigma_i\, u_i\, v_i^H$ be the singular value decomposition of $A$, and define $Q = UV^H = UIV^H$ Then $\sigma_1(Q) = 1$ by construction, and

$$\langle Q, A\rangle = \langle UV^H, U\Sigma V^H\rangle = Tr(VU^H U\Sigma V^H) = Tr(V^H VU^H U\Sigma) = Tr(\Sigma) = \sum \sigma_i$$

(Note our use of the identity $Tr(ABC) = Tr(CAB)$; this is always true when both multiplications are well-posed.) So we have established that $\sup_{\sigma_1(Q)\leq 1} \langle Q, A\rangle \geq \sum_i \sigma_i\,(A)$ . Now let's prove the other direction:

$$\sup_{\sigma_1(Q)\leq 1} \langle Q, A\rangle = \sup_{\sigma_1(Q)\leq 1} Tr(Q^H U\Sigma V^H) = \sup_{\sigma_1(Q)\leq 1} Tr(V^H Q^H U\Sigma) = \sup_{\sigma_1(Q)\leq 1} \langle UQV^H, \Sigma\rangle =$$

$$\sup_{\sigma_1(Q)\leq 1} \sum_{i=1}^{n} \sigma_i\,(UQV^H)_{ii} = \sup_{\sigma_1(Q)\leq 1} \sum_{i=1}^{n} \sigma_i u_i QV_i^H \leq \sup_{\sigma_1(Q)\leq 1} \sum_{i=1}^{n} \sigma_i\,\sigma_{max}\,(Q) = \sum_{i=1}^{n} \sigma_i$$

We have proven both the $\leq$ and $\geq$ cases, so equality is confirmed. Now we have

$$\|A+B\| = \sup_{V:\sigma_{max}(V)\leq 1} \langle V, A+B\rangle = \sup_{V:\sigma_{max}(V)\leq 1} \langle V, A\rangle + \sup_{V:\sigma_{max}(V)\leq 1} \langle V, B\rangle = \|A\| + \|B\|$$

So, weighted nuclear norm is convex because it is a norm. Now, we employ the definition of Euclidean projection onto sets in order to prove there is a unique minimizer for the distance function $J(\tau)$.

Convexity for distance function: we wish to prove that $dist(., \tau\, \partial\|.\|_*^w)$ has unique minimizer. To do that, we first use the definition of Euclidean projection onto close convex sets. Then we prove that the distance function for a convex set is convex and continuous.

Definition 3 (Euclidean projection onto sets): The Euclidean projection onto the set L is the map

$$\pi_L : \mathbb{R}^d \to L \quad \text{where} \quad \pi_L(x) := arg\,min\,\{\|x-y\| : y \in L\}$$

where $L \subset \mathbb{R}^d$ is a close convex set.

We proved that nuclear norm is convex. It is showed that the function $dist(., L)$ is convex [9, page 34] where L is convex. Therefore, $dist(., \|.\|_*)$ is convex. The next step is to prove $\pi_L$ is continuous.

Continuity for distance function: It is shown that $\pi_L$ and $\mathrm{I} - \pi_L$ is not increasing with respect to the Euclidean norm [9, page 340] and we have

$$\parallel \pi_L(x) - \pi_L(y)\parallel \leq \|x-y\| \text{ and } \parallel (I-\pi_L)(x) - (I-\pi_L)(y)\parallel \leq \|x-y\| \text{ for all } x,y \in \mathbb{R}^d \quad (9)$$

The inequality (9) indicates that $\pi_L$ is continuous. Using Lipschitz continuity, we can conclude that distance function $dis(., L)$ is continuous, i.e.

$$|dis(x, L) - dis(y, L)| \leq \|x - y\|$$

We have shown that function distance $dis(., L)$, where L is convex, is convex and continuous. So, this function is differentiable. By putting $\|.\|_*$ in the distance function instead of L, we achieve that the distance function is differentiable and, hence, there is a unique minimizer for it and we are done.

## 3. Discussion

In this paper, the weighted nuclear norm minimization was studied using convex geometry. In non-descending order of the weights, it was demonstrated that weighted nuclear norm is convex and has a unique optimizer. We also calculated the statistical dimension of descent cone for nuclear norm where phase transition occurs. This calculation showed the least number of measurements for the approximation of the low-rank matrix.

Seyedroohollah Hosseini was born in Tehran, Iran in 1989. He received his bachelor degree in 2011 from the University of Qom and his master degree in 2014 from Allameh Tabataba'i University, both in Computer Science. His research interests include convex optimization and compressed sensing.